\documentclass[12pt]{article}
\usepackage{amsthm,amsmath,amsfonts,amssymb}
\usepackage{graphicx}
\usepackage[affil-it]{authblk}
\usepackage{algorithm}
\usepackage{algorithmic}

\usepackage{multirow, booktabs}
\usepackage{subcaption}
\usepackage{bbding}
\usepackage{newpxtext}
\usepackage{setspace}  
\usepackage{natbib}
\usepackage{url} 

\newcommand{\blind}{0}

\theoremstyle{plain}
\newtheorem{theorem}{Theorem}

\newtheorem{proposition}{Proposition}

\theoremstyle{remark}
\newtheorem{remark}{Remark}
\newtheorem{definition}{Definition}
\newcommand\df{\mathrm{df}}

\newcommand\sdf{\mathrm{sdf}}
\newcommand\msdf{\mathrm{msdf}}
\newcommand\AIC{\mathrm{AIC}}
\newcommand\BIC{\mathrm{BIC}}

\newcommand\GCV{\mathrm{GCV}}
\newcommand\lasso{\mathrm{lasso}}
\newcommand\bestsubset{\mathrm{subset}}
\newcommand\cubic{\mathrm{cubic}}
\newcommand\mono{\mathrm{mono}}
\newcommand\smooth{\mathrm{smooth}}

\newcommand\RSS{\mathrm{RSS}}
\newcommand\MSE{\mathrm{MSE}}
\newcommand\Test{\mathrm{Test}}
\newcommand\ave{\mathrm{ave}}
\newcommand\IR{\mathrm{I\!R}}
\newcommand\cA{{\mathcal A}}

\newcommand\bfB{{\mathbf B}}
\newcommand\bfG{{\mathbf G}}
\newcommand\bfX{{\mathbf X}}
\newcommand\bB{{\mathbf B}}
\newcommand\bG{{\mathbf G}}
\newcommand\bI{{\mathbf I}}
\newcommand\bS{{\mathbf S}}
\newcommand\bX{{\mathbf X}}
\newcommand\bbE{{\mathbb E}}

\newcommand\bfy{{\mathbf y}}

\newcommand{\bOmega}{\boldsymbol{\Omega}}

\newcommand{\one}{\boldsymbol{1}}
\newcommand\zero{\boldsymbol{0}}

\DeclareMathOperator*{\argmin}{arg\,min}
\DeclareMathOperator*{\Cov}{Cov}
\DeclareMathOperator*{\Var}{Var}
\DeclareMathOperator*{\trace}{tr}
\DeclareMathOperator*{\rank}{rank}
\DeclareMathOperator*{\diag}{diag}
\DeclareMathOperator*{\sign}{sign}

\newcommand\textred[1]{{#1}}
\newcommand{\parencite}[1]{\citep{#1}}
\newcommand{\textcite}[1]{\citet{#1}}
\newcommand{\printbibliography}{\bibliography{bibtex-DoF}}
\usepackage[colorlinks,citecolor=blue,urlcolor=blue]{hyperref}
\newcommand{\supp}{\href{supp-jcgs.pdf}{Supplementary Material}}
\begin{document}

\def\spacingset#1{\renewcommand{\baselinestretch}%
{#1}\small\normalsize} \spacingset{1}

\date{}
\if0\blind
{
  \title{\bf Degrees of Freedom: Search Cost and Self-consistency}
\author[1,2]{Lijun Wang%
  \thanks{Email: \texttt{ljwang@link.cuhk.edu.hk}, \texttt{lijun.wang@yale.edu}}
}
\affil[1]{Department of Statistics, The Chinese University of Hong Kong, Hong Kong SAR, China}
\affil[2]{Department of Biostatistics, Yale University, New Haven, Connecticut, USA}
\author[2]{Hongyu Zhao%
\thanks{Email: \texttt{hongyu.zhao@yale.edu}}}
\author[1]{Xiaodan Fan%
  \thanks{Email: \texttt{xfan@cuhk.edu.hk}}
}
  \maketitle
} \fi

\if1\blind
{
  \bigskip
  \bigskip
  \bigskip
  \begin{center}
    {\LARGE\bf Degrees of Freedom: Search Cost and Self-consistency}
\end{center}
  \medskip
} \fi

\bigskip
\begin{abstract}
Model degrees of freedom ($\df$) is a fundamental concept in statistics because it quantifies the flexibility of a fitting procedure and is indispensable in model selection. To investigate the gap between $\df$ and the number of independent variables in the fitting procedure, \textcite{tibshiraniDegreesFreedomModel2015} introduced the \emph{search degrees of freedom} ($\sdf$) concept to account for the search cost during model selection. However, this definition has two limitations: it does not consider fitting procedures in augmented spaces and does not use the same fitting procedure for $\sdf$ and $\df$. 
We propose a \emph{modified search degrees of freedom} ($\msdf$) to directly account for the cost of searching in either original or augmented spaces. 
We check this definition for various fitting procedures, including classical linear regressions, spline methods, adaptive regressions (the best subset and the lasso), regression trees, and multivariate adaptive regression splines (MARS). In many scenarios when $\sdf$ is applicable, $\msdf$ reduces to $\sdf$. However, for certain procedures like the lasso, $\msdf$ offers a fresh perspective on search costs. For some complex procedures like MARS, the $\df$ has been pre-determined during model fitting, but the $\df$ of the final fitted procedure might differ from the pre-determined one. To investigate this discrepancy, we introduce the concepts of \emph{nominal} $\df$ and \emph{actual} $\df$, and define the property of \emph{self-consistency}, which occurs when there is no gap between these two $\df$'s. We propose a correction procedure for MARS to align these two $\df$'s, demonstrating improved fitting performance through extensive simulations and two real data applications. 

\end{abstract}

\noindent%
{\it Keywords:}  Degrees of Freedom, Model Selection, Splines, Lasso, Tree, Multivariate Adaptive Regression Splines
\vfill

\newpage
\spacingset{1.6} 

\section{Introduction}\label{sec:df_intro}

Suppose that we have observations $\{(x_i, y_i)\}_{i=1}^n, x_i \in \IR^p, y_i\in\IR$ 
from an unknown probability distribution $P(X, Y)$. Consider the estimator $\hat \mu_\lambda$ for the conditional expectation function $\mu(x)=\bbE (Y\mid X=x)$, where  the subscript in $\hat \mu_\lambda$ indicates that the estimator depends on a tuning parameter $\lambda\in\Lambda$. Determining the tuning parameter $\lambda$ is a typical task in model selection. The degrees of freedom ($\df$) plays an important role in model selection criteria, such as Akaike's information criterion (AIC),
\begin{equation}
\AIC(\lambda) = n\log\sum_{i=1}^n(y_i-\hat \mu_\lambda(x_i))^2 + 2\df_\lambda\,,    
\label{eq:aic}
\end{equation}
Bayesian information criterion (BIC),
\begin{equation}
\BIC(\lambda) = n\log\sum_{i=1}^n(y_i-\hat\mu_\lambda(x_i))^2 + \df_\lambda\log n\,,    
\label{eq:bic}
\end{equation}
and generalized cross-validation (GCV),
\begin{equation}
\GCV(\lambda)=\frac{\sum_{i=1}^n(y_i-\hat\mu_\lambda(x_i))^2}{(1-\df_\lambda/n)^2}\,, 
\label{eq:gcv}
\end{equation}
where the subscript in $\df_\lambda$ indicates that $\df$ might also depend on $\lambda$. 

\subsection{Definition of $\df$}

The concept of degrees of freedom has been widely used in many fields, and there might be ambiguity and confusion without background (\cite{goodWhatAreDegrees1973}; \cite{pandeyWhatAreDegrees2008}). In hypothesis testing scenarios, the degrees of freedom always refers to the degrees of freedom of the distribution of the test statistic under the null hypothesis, such as $t$-distribution in $t$-test, and chi-squared distribution in the Wald test. In model selection, the degrees of freedom is usually termed as the effective number of parameters in a model fitting procedure. Throughout this paper, we focus on the (model) degrees of freedom in model selection.
For linear regressions, the number of \emph{free} (linearly independent) parameters is what is meant by model degrees of freedom \parencite{hastieElementsStatisticalLearning2009}. However, there are many situations where we cannot count the number of free parameters, such as
\begin{itemize}
  \item ridge regression (Section~\ref{sec:linear_smoother}): although all $p$ coefficients are non-zero, they are fitted in a restricted fashion controlled by the penalty parameter $\lambda$.
  \item subset regression (Section~\ref{sec:adaptive}): if the subset of $k$ features is pre-specified in advance to the training data, then the number of free parameters is exactly the size of the subset, i.e., $k$; but if we carry out a best subset selection procedure to determine the optimal set of $k$ predictors, we actually use more than $k$ degrees of freedom.
\end{itemize}
To overcome those exceptions when simply counting the number of free parameters, the degrees of freedom has been defined to measure the \emph{effective} number of parameters (\cite{efronHowBiasedApparent1986}; \cite{hastieGeneralizedAdditiveModels1990}).
Suppose the observations $\{y_i\}_{i=1}^n$ are uncorrelated and have constant variance $\sigma^2$,
\begin{equation}
\bfy = \mu + \epsilon, \quad \bbE(\epsilon) = \zero_n, \quad\Cov(\epsilon) = \sigma^2\bI\,,\label{eq:df_y}
\end{equation}
where $\mu\in\IR^n$ is some fixed, true mean parameter of interest and $\bfy$ is the stacked vector of observations $\{y_i\}_{i=1}^n$. For a fitting method $\hat\mu$, denote the fitted vector as $\hat \mu(\bfy)$. The degrees of freedom of the fitting function $\hat\mu$, characterized by the fitted vector $\hat\mu(\bfy)$, has been defined as
\begin{equation}
  \df(\hat \mu) \triangleq \frac{1}{\sigma^2}\sum_{i=1}^n\Cov([\hat \mu (\bfy)]_i, y_i)\triangleq \frac{1}{\sigma^2}\sum_{i=1}^n\Cov(\hat \mu_i, y_i)\,,
  \label{eq:def_df}
\end{equation}
where for simplicity we use $\hat\mu$ both to refer to the fitted vector $\hat\mu(\bfy)$, and to the fitting function $\hat\mu:\IR^n\rightarrow\IR^n$ itself by slightly abusing notation.
Take the simple constant model as an example, $\hat \mu(\bfy) = \bar y\one$. It is easy to show that
\begin{align*}
\Cov(\hat \mu_i, y_i) &= \Cov(\bar y, y_i) = \Cov\left(\frac 1n \sum_{j=1}^n y_j, y_i\right)=\frac 1n \Cov(y_i, y_i) = \frac 1n\sigma^2\,,
\end{align*}
and hence
$$
\df(\hat\mu) = \frac{1}{\sigma^2}\sum_{i=1}^n\Cov(\hat\mu_i,y_i) = 1\,.
$$
The degrees of freedom based on the definition \eqref{eq:def_df} has been explored in \textcite{yeMeasuringCorrectingEffects1998} and \textcite{efronEstimationPredictionError2004}. \textcite{yeMeasuringCorrectingEffects1998} extended the degrees of freedom for linear models to complex modeling procedures, and presented the definition \eqref{eq:def_df} as \emph{generalized degrees of freedom}. \textcite{efronEstimationPredictionError2004} studied the covariance penalties, including Mallow's $C_p$, AIC, and Stein's unbiased risk estimate (SURE), all of which incorporate the degrees of freedom \eqref{eq:def_df}. 


\subsection{Search Cost}
\textred{As we mentioned, for adaptive procedures such as the best subset regression, the degrees of freedom exceeds the number of selected features. Then a natural follow-up question is: how many effective parameters does it \emph{cost} to search through the model space? Specifically, \emph{search cost} refers to the additional complexity introduced by the process of searching for the best model, beyond merely fitting a fixed model.
To investigate such \emph{search cost}, \textcite{tibshiraniDegreesFreedomModel2015} defined the \emph{search degrees of freedom} (sdf) for fitting procedure $\hat\mu$ as the gap between the degrees of freedom and the number of free parameters,}
\begin{equation}
\sdf(\hat\mu) = \df(\tilde \mu) - \bbE[\rank(\bX_\cA)]\,,    
\label{eq:sdf}
\end{equation}
where $\bfX$ is an $n\times p$ matrix with $x_i\in \IR^p$ in its $i$-th row, $\cA\subseteq\{1,\ldots,p\}$ is the selected active variable set, and $\tilde \mu$ is the least squares fit on the active set $\cA$. \textred{The above expectation assumes that $\bfX$ is fixed, and is taken over $\bfy$.} However, the definition is limited. Firstly, it does not establish the relationship between the \emph{search degrees of freedom} and the \emph{degrees of freedom} for the same fitting procedure $\hat\mu$. Instead, it introduces another fitting procedure $\tilde \mu$. Although $\hat\mu$ and $\tilde \mu$ can be identical when $\hat\mu$ is the best subset regression, they are usually not the same. For example, when we consider the search degrees of freedom for the lasso fit $\hat\mu$, \textcite{tibshiraniDegreesFreedomModel2015}'s definition needs first to consider the search degrees of freedom of the relaxed lasso fit $\tilde\mu$ \parencite{meinshausenRelaxedLasso2007}, which performs least squares with variables selected by the lasso.


The definition for \emph{search degrees of freedom} requires the active variable set $\cA$ of $X$. But many fitting procedures would augment or replace the input $X$ with transformations of $X$, denoted by $(h_1(X), \ldots, h_M(X))$, then the fitting procedures would be applied in this new space of derived input features. For example, the spline methods would replace the univariate $X$ with its basis expansion; the tree-based methods would consider the partition of regions, where a region can be represented by a transformation on $X$, e.g., $h_m(X) = I(L_m\le X_k< U_m)$ defines a region using the $k$-th component of $X$ with two constants $L_m, U_m$.

To overcome the above two limitations, we propose a modified definition for the \emph{search degrees of freedom}.
\begin{definition}[Modified Search Degrees of Freedom]
If the fit of a model can be written as $\hat \mu = \bS(\bfy)\bfy$, where $\bS$ might depend on $\bfy$, the \emph{modified search degrees of freedom} (msdf) is defined as
$$
\msdf(\hat\mu) = \df(\hat\mu) - \bbE[\trace(\bS(\bfy))]\,,
$$
\textred{where the expectation assumes the predictor $\bfX$ is fixed, and is taken over $\bfy$.}
\end{definition}

The \emph{search degrees of freedom} can be viewed as a special case of the \emph{modified search degrees of freedom}.
\begin{proposition}\label{prop:sdf_eq_msdf}
    If $\hat\mu$ is a composition of two steps: picking active set $\cA$ by variable selection and performing the least squares fit on $\cA$, then
    $\sdf(\hat\mu) = \msdf(\hat\mu)\,.$
\end{proposition}
    The best subset regression and the relaxed lasso are two typical examples for Proposition~\ref{prop:sdf_eq_msdf}.
    For other general fitting procedures, such as the linear smoothers (Section~\ref{sec:linear_smoother}), the lasso fit (Section~\ref{sec:df_lasso}), and tree methods (Section~\ref{sec:tree}), we can always find the linear operator $\bS(\bfy)$, then $\msdf(\hat\mu)$ would be naturally interpreted as the search cost for constructing the matrix $\bS(\bfy)$.
\subsection{Self-consistency}
In some complex fitting procedures, like the multivariate adaptive regression splines (MARS) in Section~\ref{sec:mars}, we cannot determine the search cost (and hence the degrees of freedom) before model selection, but we still need the degrees of freedom to construct one criterion (e.g., GCV) to determine the tuning parameter. Generally, we call the degrees of freedom, which needs to be pre-determined in the aforementioned criteria in Equations~\eqref{eq:aic}-\eqref{eq:gcv}, as the \emph{nominal degrees of freedom}. 
Let $\lambda\in \Lambda$ be the tuning parameter for the fitting approach $\hat\mu_\lambda$.  
After model selection with some criterion, we can obtain a particular parameter, say $\lambda^\star$, and then adopt Equation~\eqref{eq:def_df} to evaluate the degrees of freedom $\df(\hat\mu_{\lambda^\star})$, which is referred to as the \emph{actual degrees of freedom}. 
Inspired by \textcite{hastiePrincipalCurves1989}'s self-consistency concept for principal curves, we define a self-consistency property for the degrees of freedom.
 \begin{definition}[Self-consistency]\label{def:self-consistency}
    A fitting procedure $\hat\mu_\lambda$ is called \emph{self-consistent} if the \emph{actual degrees of freedom} equals to the \emph{nominal degrees of freedom} (ndf) for some $(\lambda, d)$,
    \begin{equation}
            \df(\hat\mu_{\lambda} \mid \mathrm{ndf}(\hat\mu_{\lambda}) = d) = d\,.
            \label{eq:self-consistency0}
    \end{equation}
    \textred{
    Let $\mathrm{ndf}^{-1}(d) \triangleq \{\lambda: \mathrm{ndf}(\hat\mu_\lambda) = d\}$, then the condition \eqref{eq:self-consistency0} can also be written as}%
    \begin{equation}
    \textred{\df(\hat\mu_{\lambda}\mid \lambda \in \mathrm{ndf}^{-1}(d)) = d\,.    }
    \label{eq:self-consistency}
    \end{equation}
\end{definition}%
\textred{%
\begin{remark}
Given a nominal degrees of freedom $d$, there might be multiple $\lambda$ such that $\mathrm{ndf}(\hat\mu_{\lambda}) = d$, so $\mathrm{ndf}^{-1}(d)$ is a set instead of a single value. In other words, there might exist multiple pairs $(\lambda, d)$ that fulfill the self-consistency property.
\end{remark}
}

If we can calculate the degrees of freedom before model selection, just set the nominal degrees of freedom as $\df(\hat\mu_\lambda)$, then the self-consistency property would be automatically satisfied. However, for approaches like MARS, the \emph{nominal degrees of freedom} is more like a hypothesis instead of a derivation from the formula of the degrees of freedom, so self-consistency generally cannot hold. \textred{Consequently, MARS often uses an incorrect GCV when the \emph{nominal degrees of freedom} does not equal the \emph{actual degrees of freedom}. Pursuing self-consistency helps correct the misspecified degrees of freedom in GCV. We propose a correcting procedure to fulfill the self-consistency property in Section~\ref{sec:mars}. Based on simulations and real data analyses, we show that the corrected MARS significantly improves model performance compared to the default MARS.}

Since it is usually hard to calculate the theoretical degrees of freedom by Equation~\eqref{eq:def_df}, we resort to a Monte Carlo method, summarized in Algorithm \ref{alg:dfhat}, to approximate the degrees of freedom, which is termed as \emph{empirical degrees of freedom}.
\begin{algorithm}[H]
  \caption{Empirical Degrees of Freedom}
  \label{alg:dfhat}
  \begin{algorithmic}[1]
    \REQUIRE Sample size $n$; number of Monte Carlo repetitions $m$.
    \REQUIRE (Optional) Design matrix $\bX$ of size $n\times p$, and coefficient $\beta$. 
    \REQUIRE Truth vector $\mu$. If both $\bX$ and $\beta$ are given, $\mu = \bX\beta$; otherwise $\mu = \zero_n$.
    \STATE \COMMENT{\texttt{Repeat data generation for $m$ times.}}
    \FOR{$j=1$ \TO $m$}
    \STATE \COMMENT{\texttt{Generate $n$ observations independently.}}
    \FOR{$i=1$ \TO $n$}
    \STATE simulate $y_{ij}\sim N(\mu_i, 1)$.
    \ENDFOR
    \ENDFOR
    \STATE\COMMENT{\texttt{Conduct fitting for each repetition}}
    \FOR{$j=1$ \TO $m$}
    \STATE Fit the $j$-th column vector $y_{\cdot j}$ to yield $\hat\mu^{(j)}$.
    \ENDFOR
    \FOR{$i=1$ \TO $n$}
    \STATE Let $\boldsymbol{\hat\mu_i} = [\hat\mu_i^{(1)},\ldots, \hat\mu_i^{(m)}]$, $y_{i\cdot} = [y_{i1}, \ldots, y_{im}]$.
    \STATE Calculate the sample covariance for each observation:
    $$
    c_i = \widehat\Cov(\boldsymbol{\hat\mu_i}, y_{i\cdot})\,.
    $$
    \ENDFOR
    \STATE The empirical degrees of freedom is $\widehat\df=\sum_{i=1}^nc_i$\,.
  \end{algorithmic}
\end{algorithm}






%
%

\subsection{Organization}

\begin{table}[H]
    \centering
    \caption{Paper Organization. $n_\mathrm{coef}$ is the number of free parameters. The check and cross symbols indicate where there exist the search degrees of freedom (sdf) and the modified search degrees of freedom (msdf). The comparisons between $n_\mathrm{coef}$ and $\df$ do not consider trivial (or reduced) cases, such as the penalty parameter $\lambda=0$ in ridge regressions.}
    \label{tab:org}
    \begin{tabular}{cccc}
         &  $n_{\mathrm{coef}} \bigcirc \df$ & sdf & msdf \\
      Ridge (Section~\ref{sec:linear_reg}) & > &  \CheckmarkBold & \CheckmarkBold\\
      Smoothing Spines (Section~\ref{sec:neq_constraint}) & > & \XSolidBrush & \CheckmarkBold\\
      Monotone Cubic Splines (Section~\ref{sec:neq_constraint}) & > & \XSolidBrush & \CheckmarkBold\\
      Cubic Spines (Section~\ref{sec:neq_constraint}) & = & \XSolidBrush & \CheckmarkBold\\
      Ordinary Least Squares (Section~\ref{sec:linear_reg}) & = &  \CheckmarkBold & \CheckmarkBold\\
      Lasso (Section~\ref{sec:df_lasso})   & = & \CheckmarkBold&\CheckmarkBold \\
        Best Subset (Section~\ref{sec:subset}) & < & \CheckmarkBold&\CheckmarkBold\\
      Tree (Section~\ref{sec:regtree}) & < & \XSolidBrush & \CheckmarkBold\\
      MARS (Section~\ref{sec:mars}) & < & \XSolidBrush & \CheckmarkBold
    \end{tabular}
\end{table}

The remaining of the paper is organized as follows, which is also summarized in Table~\ref{tab:org}.
Section~\ref{sec:linear_smoother} discusses regularization and constrained methods, such as ridge regressions (Section \ref{sec:linear_reg}), smoothing splines and monotone cubic splines (Section \ref{sec:neq_constraint}), each of whose degrees of freedom tends to be smaller than the number of free parameters. 
Section~\ref{sec:adaptive} investigates methods with variable selection, such as the best subset selection (Section \ref{sec:subset}), whose degrees of freedom tends to be larger than the number of free parameters, and the lasso (Section~\ref{sec:df_lasso}), whose degrees of freedom is exactly the number of selected variables. We take another perspective to study the degrees of freedom of the lasso, and show that it also exhibits a nonzero search cost based on our \emph{modified search degrees of freedom} definition. Section \ref{sec:tree} will discuss the tree-based and tree-like methods. The tree-based method refers to the regression tree (Section \ref{sec:regtree}), which is shown to have a large search cost. MARS (Section~\ref{sec:mars}) is viewed as a tree-like method. We elaborate on its violation of self-consistency and the correction procedure and demonstrate the improvement of the corrected MARS using simulations and real data applications. Limitations and potential future work are discussed in Section \ref{sec:df_future_work}.

\section{Linear Smoothers}\label{sec:linear_smoother}

If the fitting procedure $\hat\mu$ is a linear smoother, then there exists a smooth matrix $\bS$ such that $\hat\mu = \bS\bfy$, where $\bS$ does not depend on $\bfy$. 
Since $\hat\mu_i = \sum_{j=1}^n\bS_{ij}y_j$, then by the definition \eqref{eq:def_df}, the degrees of freedom turns out to be:
\begin{align*}
  \frac{1}{\sigma^2}\sum_{i=1}^n\Cov(\hat\mu_i, y_i) &= \frac{1}{\sigma^2}\sum_{i=1}^n\Cov\left(\sum_{j=1}^n\bS_{ij}y_j, y_i\right) \\
  &=\frac{1}{\sigma^2}\sum_{i=1}^n\sum_{j=1}^n\bS_{ij}\Cov(y_j, y_i)=\frac{1}{\sigma^2}\sum_{i=1}^n\bS_{ii}\sigma^2=\trace(\bS)\,.
\end{align*}
It follows that the \emph{modified search degrees of freedom} is 
$$
\msdf(\hat\mu) = \trace(\bS) - \bbE[\trace(\bS)] = 0\,,
$$
which implies that the linear smoother does not need extra effort to construct the smooth matrix $\bS$. 
Once $\bS$ is given, we can easily evaluate the degrees of freedom $\df$ and plug it into model selection criteria, so the self-consistency property would be automatically satisfied.
\subsection{Linear Regressions}\label{sec:linear_reg}
We present several well-known linear regressions, which are special cases of linear smoothers.
\begin{itemize}
  \item ordinary least squares: $\hat \mu = \bX(\bX^\top\bX)^{-1}\bX^\top\bfy$, then $\df = \trace(\bX(\bX^\top\bX)^{-1}\bX^\top) = p$, where $\bX$ is the $n\times p$ design matrix and assumed to have full column rank.
  \item ridge regression: $\hat\mu = \bX(\bX^\top\bX+\lambda\bI)^{-1}\bX^\top\bfy$, then $\df = \trace(\bX(\bX^\top\bX+\lambda \bI)^{-1}\bX^\top) = \sum_{j=1}^p\frac{d_j^2}{d_j^2+\lambda}$, where the $d_j$'s are the singular values of $\bX$.
  \item $k$-nearest-neighbor averaging: at each point $x$, the fitting is the average of the responses of its neighbors, that is,
  $\hat\mu(x) = \mathrm{Ave}(y_i\mid x_i\in N_k(x))$\,,
  where $N_k(x)$ is the neighborhood containing the $k$ points cloest to $x$. We can write it in matrix form,
  $$
  \hat \mu = \frac{1}{k}\begin{bmatrix}
  1 & * & \cdots & *\\
  * & 1 & \cdots & *\\
  \vdots & \vdots & \ddots & *\\
  * & * & \cdots & 1
\end{bmatrix}
\bfy\triangleq \bS\bfy\,,
  $$
  where the $*$  symbol denotes unknown (but uninterested) values, then the degrees of freedom would be $\df = \trace(\bS) = n/k$.

\end{itemize}
\textcite{hodgesCountingDegreesFreedom2001} also derived the degrees of freedom for linear hierarchical and random-effects models.



\subsection{Spline Methods}\label{sec:neq_constraint}
\subsubsection{Cubic Splines and Smoothing Splines}
In the spline fitting, we want to find some function $f$ by minimizing
\begin{equation}
\sum_{i=1}^n(y_i-f(x_i))^2 + \lambda\int \{f''(t)\}^2dt.\label{eq:df_spline}
\end{equation}
A widely-used approach is to take cubic B-splines as the basis for the solution,
$f(x) = \sum_{j=1}^J\gamma_jB_j(x)\,,$
where $B_j,j=1,\ldots,J$ are basis functions, and $\gamma_j,j=1,\ldots,J$ are the coefficients. Stack the observations $y_i$, coefficients $\gamma_i$ into the vectors $\bfy,\gamma$, respectively, and define $\{\bfB\}_{ij}=B_j(x_i),i=1,\ldots,n,j=1,\ldots,J$ as the evaluation of the $j$-th B-spline basis at point $x_i$. 
Now problem \eqref{eq:df_spline} can be expressed in a matrix form,
\begin{equation}
  \hat\gamma^{\lambda,J} = \argmin_{\gamma} (\bfy - \bfB\gamma)^\top(\bfy - \bfB\gamma) + \lambda \gamma^\top\bOmega\gamma\label{eq:matrss_cubic_smooth_spline_df}\,,
\end{equation}
where $\{\bOmega\}_{jk}=\int B_j''(s)B_k''(s)ds$ is the penalty matrix. 
The fitting turns out to be
$$
\hat\mu^{\lambda,J} = \bfB\hat\gamma^{\lambda,J}
= \bfB(\bfB^\top\bfB+\lambda\bOmega)^{-1}\bfB^\top\bfy\triangleq \bS_\lambda \bfy\,.
$$

If $\lambda = 0$, this fitting $\hat\mu^\cubic(J)\triangleq\hat\mu^{0,J}$ reduces to a \emph{cubic spline}. When $\lambda >0$, it becomes a \emph{smoothing spline} (or \emph{natural spline}), and in that case, the number of basis functions is usually fixed, which is completely determined by the number of unique $x$'s. Specifically, if all $x$'s are unique, then $J=n+4$, in which 4 is the order of cubic splines.
As a linear smoother, the degrees of freedom is
$$
\df(\hat\mu^{\lambda,J})=\trace(\bfB(\bfB^\top\bfB+\lambda\bOmega)^{-1}\bfB^\top)\,,
$$
and particularly when $\lambda = 0$,
$
\df(\hat\mu^{0, J}) = J\,.
$

\subsubsection{Monotone Splines}
If the coefficients in Equation \eqref{eq:matrss_cubic_smooth_spline_df} are restricted to be monotone, $\gamma_1 \le \cdots \le \gamma_J$,
the resulting solution
$\hat\mu^{\lambda,J,\mono} = \bfB\hat\gamma^{\lambda, J,\mono}$
would be a monotone spline
since the increasing coefficients imply an increasing spline \parencite{wangMonotoneCubicBSplines2023c}. We consider the \emph{monotone cubic spline} $\hat\mu^{\mono,\cubic}(J)\triangleq \hat\mu^{0, J, \mono}$, and the \emph{monotone smoothing spline} $\hat\mu^{\mono,\smooth}(\lambda)\triangleq \hat\mu^{\lambda, J, \mono}$.

\textcite{chenDegreesFreedomProjection2020} studied the degrees of freedom of nonparametric estimators for least squares problems with linear constraints and/or quadratic penalties. 
We can apply their results on monotone cubic splines to obtain Proposition \ref{coro:df_mono_cubic}. 
\begin{proposition}\label{coro:df_mono_cubic}
The degrees of freedom for the monotone cubic B-spline $\hat\mu^{\mono,\cubic}(J)$ is
\begin{equation}
  \df = \bbE[U_\bfy]\,,
  \label{eq:df_cubic_mono_spl}
\end{equation}
where $U_\bfy$ (depends on $\bfy$) is the number of unique coefficients.
\end{proposition}
\begin{remark}
Although we can also derive theoretical degrees of freedom for the monotone smoothing splines $\hat\mu^{\mono,\smooth}(\lambda)$ by applying \textcite{chenDegreesFreedomProjection2020}'s Theorem, it is much more complicated and we cannot obtain a simpler formula like Proposition \ref{coro:df_mono_cubic}. Furthermore, the derived formula is not numerical-stable due to the matrix inversion operation.    
\end{remark}

\textcite{wangMonotoneCubicBSplines2023c} shows that we can also write the solutions for monotone splines in ``linear smoother'' form. Particularly, for monotone cubic splines, there exists a matrix $\bfG$ of size $g_\bfy\times J$ such that
$$
\hat\mu^{\mono,\cubic}(J) = \bfB\bG^\top(\bG\bB^\top\bB\bG^\top)^{-1}\bG\bB^\top\bfy\triangleq \bS_\bfy\bfy\,,
$$
\textred{
where
\begin{equation*}
    \bG = \begin{bmatrix}
\bI_{k_1-1} & & &&&\\
& \one^T_{k_2-k_1+1} & &&&\\
& & \ddots &&&\\
 &  &  & \bI_{k_{m-1}-k_{m-2}-1} &&\\
 &  & &  & \one^T_{k_{m}-k_{m-1}+1} &\\
 &  & &  & & \bI_{J-k_{m}}
\end{bmatrix}\,,
\label{eq:md_g}
\end{equation*}
in which $\one$ is the all-ones vector, $\bI$ is the identity matrix and $1\le k_1\le k_2\ldots\le k_m\le J$ are indexes related to the solution (and hence depend on $\bfy$); and} $g_\bfy$ is the number of unique coefficients. Note that $\bS_\bfy$ depends on $\bfy$ since both $\bG$ and $g_\bfy$ depend on $\bfy$, so it differs from the standard linear smoother. However, if we still adopt $\trace(\bS_\bfy)$ as the degrees of freedom but take expectation over $\bfy$, then we have 
$$
\df = \bbE[\trace(\bS_\bfy)] =\bbE[g_\bfy]\,,
$$
which coincides with Proposition~\ref{coro:df_mono_cubic}.

We apply Algorithm \ref{alg:dfhat} on the aforementioned spline methods, and repeat for 100 times to obtain the average empirical degrees of freedom, together with their standard errors, which are shown in Table \ref{tab:df_splines}. The theoretical degrees of freedom for $\hat\mu^{\mono,\cubic}(J)$ are approximated by the Monte Carlo estimates of the expectation in Equation \eqref{eq:df_cubic_mono_spl}. Table \ref{tab:df_splines} shows that the differences between the empirical degrees of freedom and the theoretical results are quite small. Comparing $\hat\mu^\cubic$ to the corresponding $\hat\mu^{\mono,\cubic}$ and comparing $\hat\mu^\smooth$ to the corresponding $\hat\mu^{\mono,\smooth}$, the monotone constraint can further shrink the degrees of freedom since it forces the splines to be simpler.

\begin{table}
  \centering
    \caption[Theoretical and Empirical Degrees of Freedom for Spline Methods]{The theoretical and empirical degrees of freedom for spline methods when \textred{the sample size $n=100$ and the number of Monte Carlo repetitions $m=100$}. The empirical results are averaged over 100 simulations, with the standard error in parentheses.}
  \label{tab:df_splines}
  \begin{tabular}{lrrr}
\toprule
Method & Parameter & Theoretical & Empirical\tabularnewline
\midrule
\multirow{3}{*}{$\hat\mu^\cubic(J)$}& 5 & 5.0 & 4.95 (0.033)\tabularnewline
& 10 & 10.0 & 9.99 (0.049)\tabularnewline
& 15 & 15.0 & 15.03 (0.052)\tabularnewline
\midrule
\multirow{3}{*}{$\hat\mu^\smooth(\lambda)$}& 0.001 & 7.33 & 7.31 (0.032) \tabularnewline
& 0.01 & 4.56 & 4.60 (0.026) \tabularnewline
& 0.1 & 3.00 & 3.01 (0.021) \tabularnewline
\midrule
\multirow{3}{*}{$\hat\mu^{\mono,\cubic}(J)$}& 5 & 2.21 & 2.08 (0.021)\tabularnewline
& 10 & 2.78 & 2.80 (0.020)\tabularnewline
& 15 & 3.38 & 3.21 (0.023)\tabularnewline
\midrule
\multirow{3}{*}{$\hat\mu^{\mono,\smooth}(\lambda)$}& 0.001 & - & 2.85 (0.020)\tabularnewline
& 0.01 & - & 2.33 (0.017)\tabularnewline
& 0.1 & - & 1.90 (0.018)\tabularnewline
\bottomrule
\end{tabular}

\end{table}


\section{Adaptive Regressions}\label{sec:adaptive}

Besides the least-squares regressions and ridge regressions discussed in Section \ref{sec:linear_reg}, there are other estimators approximating the response variable using a linear combination of the predictors, such as the best subset regression and the lasso, both of which choose a subset of variables adaptively. 


\subsection{Best Subset Regression}\label{sec:subset}
The best subset selection estimator can be expressed as
$$
\hat\beta^\bestsubset \in \argmin_{\beta\in\IR^p} \Vert \bfy-\bX\beta\Vert^2 + \lambda\Vert \beta\Vert_0\,,
$$
where $\Vert\beta\Vert_0 = \sum_{j=1}^p1\{\beta_j\neq 0\}$. \textcite{tibshiraniDegreesFreedomModel2015} showed that the degrees of freedom is larger than the number of free parameters in the orthogonal case ($\bfX^\top\bfX = \bI$). 

The \emph{search degrees of freedom} defined in Equation~\eqref{eq:sdf} turns out to be
$$
\sdf(\hat\mu^\bestsubset) = \df(\hat\mu^\bestsubset)-\bbE\vert\cA^\bestsubset\vert\,.
$$
For the \emph{modified search degrees of freedom}, since the best subset regression satisfies Proposition~\ref{prop:sdf_eq_msdf}, it follows that $\msdf(\hat\mu^\bestsubset)=\sdf(\hat\mu^\bestsubset)$.

\subsection{Lasso}\label{sec:df_lasso}
The lasso \parencite{tibshiraniRegressionShrinkageSelection1996} also performs variable selection, and the estimate is
\begin{equation}
\hat\beta^\lasso =\argmin_{\beta\in\IR^p} \Vert \bfy-\bX\beta\Vert^2 + \lambda\Vert\beta\Vert_1\,.\label{eq:lasso}
\end{equation}
The degrees of freedom of the lasso has been studied in \textcite{zouDegreesFreedomLasso2007} and \textcite{tibshiraniDegreesFreedomLasso2012}. They showed that its degrees of freedom would be the number of free parameters.

\textcite{tibshiraniDegreesFreedomModel2015} argued that the \emph{search degrees of freedom} of the lasso equals the one of the relaxed lasso, which refits with the selected variable set $\cA$ from the lasso. The argument might not be counterintuitive since the lasso does not have the refitting step as in the relaxed lasso, so such a \emph{search degrees of freedom} cannot account for the search effort of the lasso in the adaptive procedure.
\subsubsection{Approximate Lasso by Iterative Ridge}

We take another perspective to study the degrees of freedom of the lasso by constructing the solution with a linear operator and show that there exists a nonzero \emph{modified search degrees of freedom} for the lasso.

First of all, let us start with two general scalar functions. Consider $f(u) = \vert u\vert$, and
$g(u, v) = \vert v\vert + \frac{1}{2\vert v\vert} (u^2-v^2)$\,,
then we have
\begin{align*}
g(u, u) = f(u);\quad g(u, v) = \frac{u^2}{2\vert v\vert} + \frac{\vert v\vert}{2}\ge 2\sqrt{\frac{u^2}{2\vert v\vert} \frac{\vert v\vert}{2}}=\vert u\vert=f(u)\,.    
\end{align*}
It implies that $g(u, v)$ \emph{majorizes} $f(u)$, then the minimization of $f(u)$ can be done with the update
$u^{(k+1)} = \argmin_u g(u, u^{(k)})$, since $f(u)$ is nonincreasing on the sequence $\{u^{(k)}\}_{k=1}^\infty$,
\begin{equation*}
    \begin{split}
f(u^{(k+1)}) &= g(u^{(k+1)}, u^{(k+1)}) \le g(u^{k+1},u^{(k)})\le g(u^{(k)}, u^{(k)})=f(u^{(k)})\,.        
    \end{split}
\end{equation*}
This is well known as the majorize-minimize (MM) algorithm \parencite{langeOptimization2013}.

Note that the objective function in the lasso problem \eqref{eq:lasso} can be written as
$$
\RSS(\beta) = \Vert y-\bX\beta \Vert_2^2 + \lambda \Vert\beta\Vert_1 = \Vert y-\bX\beta \Vert_2^2 + \lambda \sum_{i=1}^p\vert\beta_j\vert\,,
$$
then one majorization can be chosen to be
$$
\RSS(\beta,\theta) =  \Vert y-\bX\beta \Vert_2^2 + \lambda \sum_{i=1}^p\vert\theta_j\vert + \lambda\sum_{i=1}^p\frac{1}{2\vert \theta_j\vert}(\beta_j^2-\theta_j^2) \,.
$$
Given current estimation $\beta^{(k)}$, the next update can be written as
\begin{align*}
\hat\beta^{(k+1)} &= \argmin_\beta\RSS(\beta, \hat\beta^{(k)})=\argmin_\beta \Vert \bfy-\bX\beta \Vert_2^2 + \lambda \beta'\Psi^{(k)}\beta\\
&=(\bX^\top\bX+\lambda \Psi^{(k)})^{-1}\bX^\top\bfy\,,
\end{align*}
where $\Psi^{(k)}$ is a diagonal matrix with elements $\left\{\frac{1}{2\vert\beta_j^{(k)}\vert}\right\}_{j=1}^p$. The update can be viewed as a generalized ridge regression, and the initialization can be taken to be the ridge estimation, $\hat\beta^{(1)} = (\bX^\top\bX+\lambda \bI)^{-1}\bX^\top\bfy\,.$
Since the lasso tends to do variable selection, some coefficients $\beta_j$ will be zero. In the above iteration, although $\beta_j$ might not be exactly zero, its inverse would approach infinity, and hence the penalty parameter for the $j$-th coefficient would approach infinity. To overcome such an issue, \textcite{vanwieringenLectureNotesRidge2021} suggested removing the $j$-th covariate from the model altogether. But based on our experiments, removing the covariates would result in a different solution, which is far from the lasso estimate. Instead, we set the coefficients near zero as a small number, say, $1.0\times 10^{-7}$.

Once the iterative ridge has converged, since the iterative ridge converges to the lasso solution, the lasso solution can be written as $\hat\beta^\lasso = (\bX^\top\bX + \lambda \Psi)^{-1}\bX^\top\bfy\,,$
where $\Psi = \diag\{1/\vert\hat\beta_1^\lasso\vert,\ldots,1/\vert\hat\beta_p^\lasso\vert\}$, then the fitted value is
$$
\hat \bfy = \bX\hat\beta^\lasso = \bX(\bX^\top\bX + \lambda \Psi)^{-1}\bX^\top\bfy\triangleq \bS\bfy\,.
$$
It follows that $\trace(\bS)$ might serve as an estimate for the degrees of freedom for the lasso based on Section \ref{sec:linear_smoother}. But Theorem \ref{prop:df_lasso} shows that it is a biased estimate, and it always underestimates. 

\begin{theorem}\label{prop:df_lasso}
The degrees of freedom for the lasso can be characterized by
$$
\df = \bbE[\delta + \trace \bS(\bfy)]\,,
\qquad
\delta = \sum_{i=1}^n\sum_{j=1}^n\left(\frac{\partial \bS_{ij}(\bfy)}{\partial y_i}y_j\right)\,,
$$
where
\begin{align*}
    \frac{\partial \bS(\bfy)}{\partial y_i} &= -\lambda \bX(\bX^\top\bX+\lambda\Psi(\bfy))^{-1}C_{\hat\beta}\cdot D_{\hat\beta}(\bX^\top\bX+\lambda\Psi(\bfy))^{-1}\bX^\top\\
C_{\hat\beta} &= \diag\left\{\frac{-\sign(\hat\beta_1)}{2\hat\beta_1^2},\ldots,\frac{-\sign(\hat\beta_p)}{2\hat\beta_p^2}\right\}\,,
D_{\hat\beta} = \diag\left\{\frac{\partial \hat\beta_1}{\partial y_i},\ldots,\frac{\partial \hat\beta_p}{\partial y_i}\right\}\,,
\end{align*}
and $\sign(x)$ is the sign function, which takes 1 if $x > 0$, -1 if $x < 0$, and zero if $x=0$, and define $0/0$ as 0 when $\hat\beta_j=0$.
\end{theorem}

By the definition of \emph{modified search degrees of freedom}, we have
$$
\msdf = \df - \bbE[\trace\bS(\bfy)] = \delta\,,
$$
which can be thought as the amount that comes from the iterations to determine (search) $\Psi$. As a comparison, \textcite{tibshiraniDegreesFreedomModel2015}'s \emph{search degrees of freedom} can only give the search cost for the relaxed lasso. 

In practice, the finite difference can be used to approximate the derivative of $\hat\beta$ with respect to $\bfy$. Then we can compare the degrees of freedom calculated from Theorem \ref{prop:df_lasso} and the number of nonzero coefficients $\vert\cA\vert$. 

\subsubsection{Examples}

Here are some simulations for comparing the solutions, degrees of freedom and GCV for the iterative ridge and the lasso.
We generate $\{x_{i1},\ldots,x_{ip}, y_i\},i=1,\ldots,p$ from
$y_i = \sum_{j=1}^p x_{ij}\beta_j + \varepsilon_i\,,$
where both $x_i$ and $\varepsilon_i$ are sampled independently from the standard Gaussian distribution. Let $\beta_1=\beta_2=\beta_3 = 1$ and $\beta_4=\cdots=\beta_p=0$ such that the signal-to-noise ratio $\Var[\bbE(Y\mid X)]/\Var(\varepsilon)$ is 3.

\begin{figure}
    \centering
    \begin{subfigure}{0.5\textwidth}
    \includegraphics[page=1,width=\textwidth]{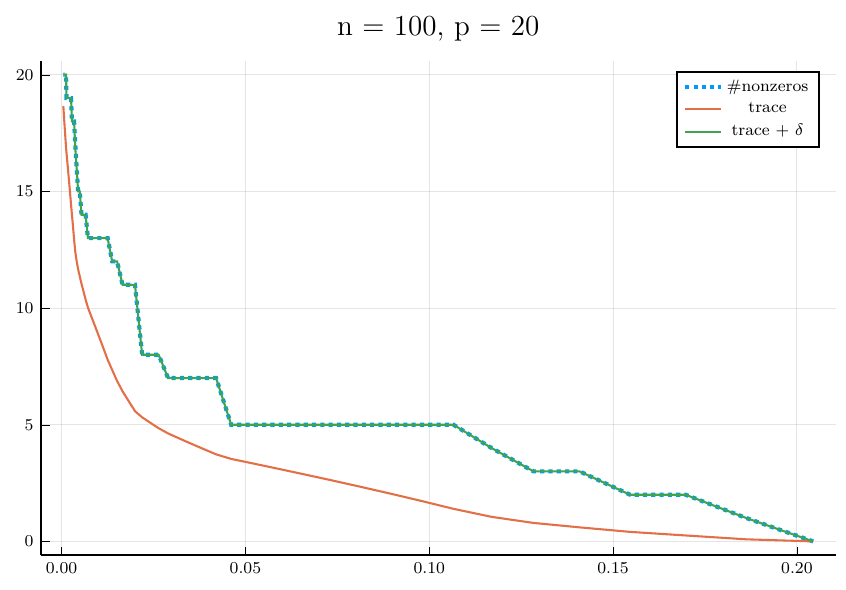}
    \end{subfigure}%
    \begin{subfigure}{0.5\textwidth}
    \includegraphics[page=2,width=\textwidth]{res/df/demo_iter_ridge_df_n100_p20.pdf}
    \end{subfigure}
    \begin{subfigure}{0.5\textwidth}
    \includegraphics[page=1,width=\textwidth]{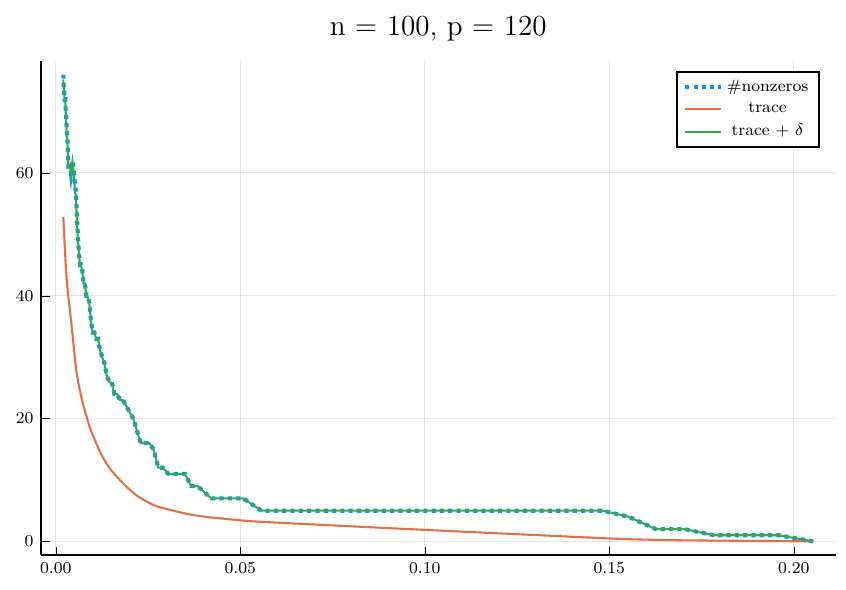}
    \end{subfigure}%
    \begin{subfigure}{0.5\textwidth}
    \includegraphics[page=2,width=\textwidth]{res/df/demo_iter_ridge_df_n100_p120.pdf}
    \end{subfigure}
    \caption[Different Estimates for Degrees of Freedom of Lasso]{(\emph{Left}) Different estimates for the degrees of freedom of the lasso. The dashed line counts the number of nonzero coefficients, and the red curve calculates $\trace(\bS)$, and the green line corrects the red curve by adding $\delta$. (\emph{Right}) Difference between the estimate $\delta + \trace(\bS)$ and the estimate by counting nonzero coefficients. }
    \label{fig:df_lasso}
\end{figure}
Figure \ref{fig:df_lasso} agrees with Theorem \ref{prop:df_lasso}, which shows that the trace of the smooth matrix via iterative ridge would always underestimate the degrees of freedom, but these two methods will coincide after adding the corrected term $\delta$.

\begin{figure}
    \begin{subfigure}{0.33\textwidth}
    \includegraphics[page=2,width=\textwidth]{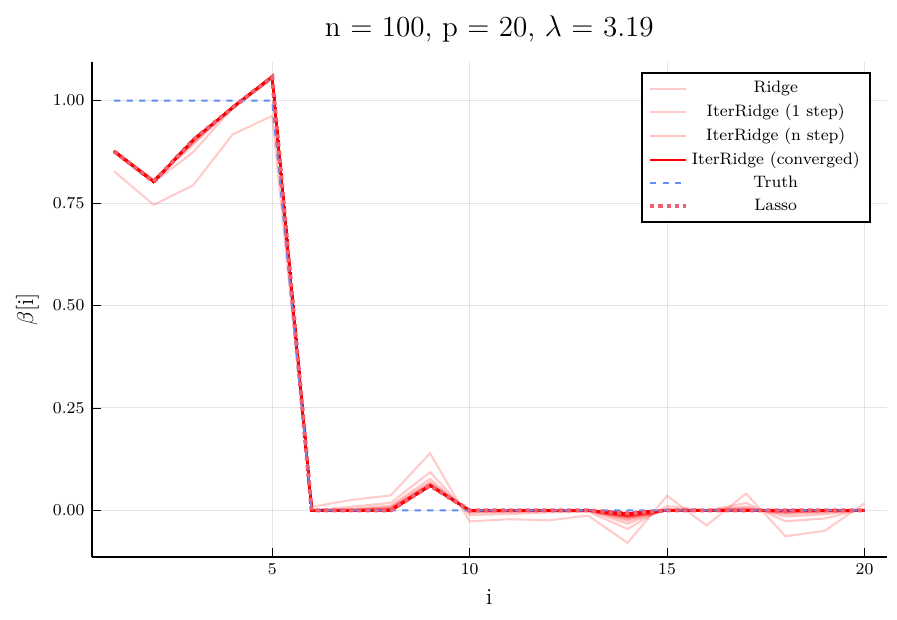}
    \end{subfigure}%
    \begin{subfigure}{0.33\textwidth}
    \includegraphics[page=3,width=\textwidth]{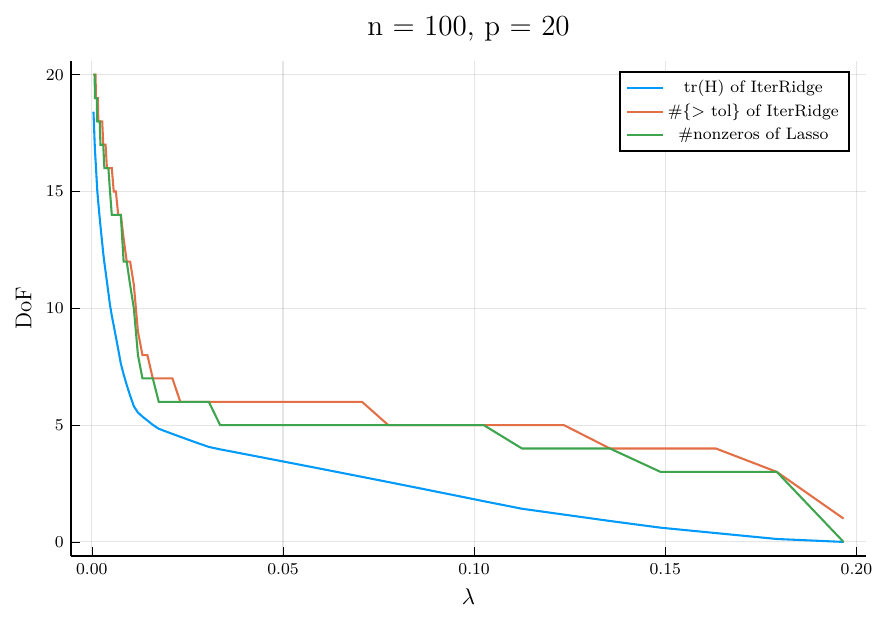}
    \end{subfigure}%
    \begin{subfigure}{0.33\textwidth}
    \includegraphics[page=4,width=\textwidth]{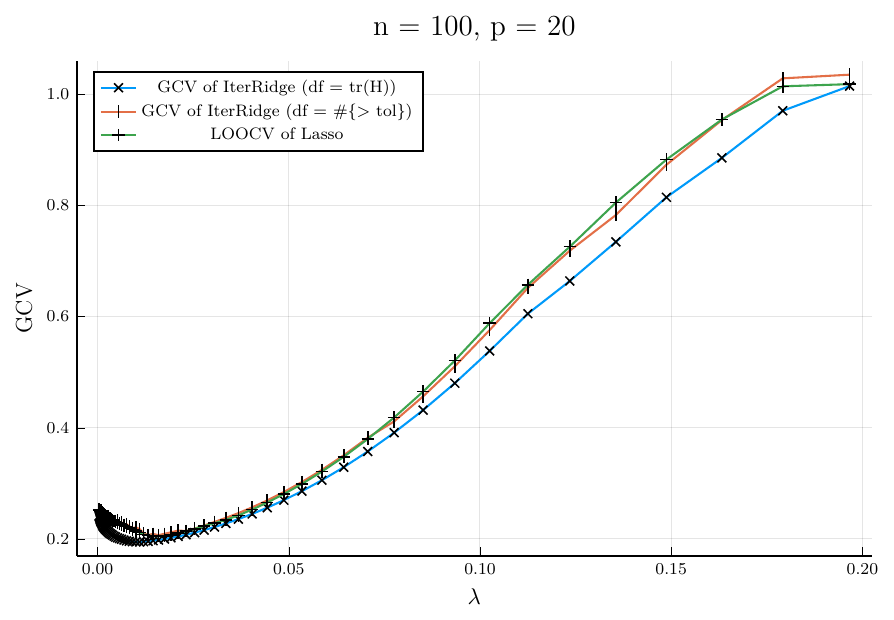}
    \end{subfigure}
    \caption[Iterative Ridge Regression when $n>p$]{Demo of iterative ridge regression when $n > p$. The left panel shows the ridge solution at each iteration, and a thicker color denotes a solution with more iterations. The middle panel shows the degrees of freedom calculated based on the trace of $\bS$ at the last iteration, and the ones calculated by counting the number of nonzero coefficients. The right panel shows the GCV calculated based on the iterative ridge and the LOOCV of the lasso.}
    \label{fig:demo_small_p}
\end{figure}

The left panel of Figure \ref{fig:demo_small_p} shows that the iterative ridge indeed has a good approximation for the lasso at each $\lambda$. The resulting GCV in the right panel of Figure \ref{fig:demo_small_p} from iterative ridge can also achieve a good approximation to the leave-one-out cross-validation (LOOCV) of the lasso, regardless of two different degrees of freedom, although the one by counting the number of (nearly) zero coefficients seems better.

\begin{figure*}
    \begin{subfigure}{0.33\textwidth}
    \includegraphics[page=2,width=\textwidth]{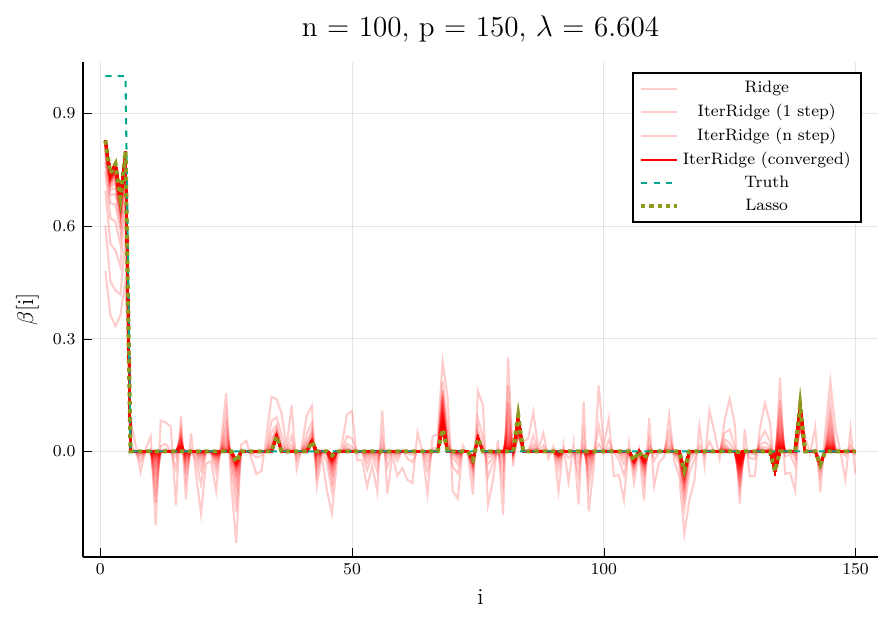}
    \end{subfigure}%
    \begin{subfigure}{0.33\textwidth}
    \includegraphics[page=3,width=\textwidth]{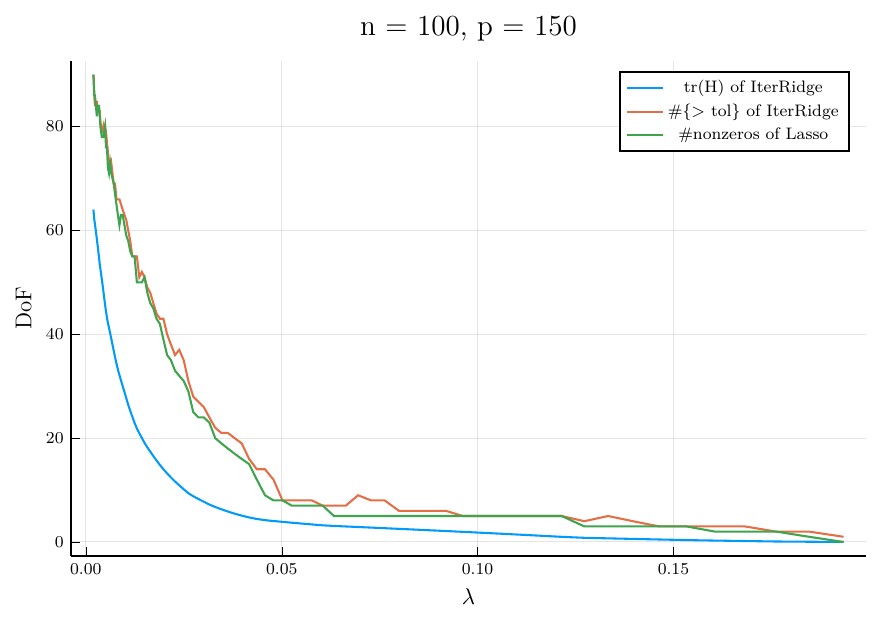}
    \end{subfigure}%
    \begin{subfigure}{0.33\textwidth}
    \includegraphics[page=4,width=\textwidth]{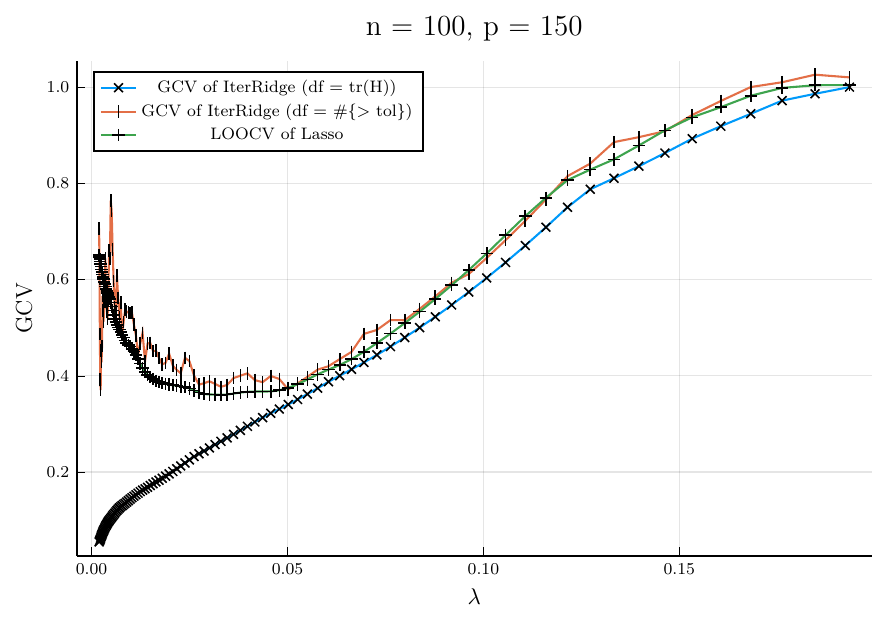}
    \end{subfigure}
    \caption[Iterative Ridge Regression when $n < p$]{Demo of iterative ridge regression when $n < p$. The left panel shows the ridge solution at each iteration, and a thicker color denotes a solution with more iterations. The middle panel shows the degrees of freedom calculated based on the trace of $\bS$ at the last iteration, and the ones calculated by counting the number of nonzero coefficients. The right panel shows the GCV calculated based on the iterative ridge and the LOOCV of the lasso.}
    \label{fig:demo_large_p}
\end{figure*}

Figure \ref{fig:demo_large_p} shows the results when $n < p$. The iterative ridge again converges to the lasso solution, and the GCV curve via the degrees of freedom by counting the (nearly) zero coefficients is close to LOOCV, but the GCV curve via the trace of $\bS$ has different behavior. The reason is that the trace underestimates the degrees of freedom when $\lambda$ is small.

\section{Tree-based and Tree-like Methods}\label{sec:tree}
\subsection{Regression Tree}\label{sec:regtree}

The phenomenon that the degrees of freedom can be larger than the number of free parameters has also been observed in the regression tree. The data consists of $p$ inputs and a response: $\{(x_i,y_i)\}_{i=1}^n$ with $x_i=(x_{i1},\ldots,x_{ip})$. 
To grow a regression tree, suppose there is a partition into $M$ regions $R_1,\ldots,R_M$, and we model the response as a constant $c_m$ in each region. The conventional criterion for the regression tree is the minimization of the sum of squares $\sum (y_i-f(x_i))^2$. It can be shown that the best $\hat c_m$ is just the average of $y_i$ in the region $R_m$, then the fitting function can be written as
\begin{align*}
\hat \mu(x) &= \sum_{m=1}^M\hat c_mI(x\in R_m) = \sum_{m=1}^M\frac{\sum_{j=1}^n y_jI(x_j\in R_m)}{\sum_{k=1}^n I(x_k\in R_m)}I(x\in R_m)\,.    
\end{align*}
Switch the summation for $m$ and $j$ in the numerator, then for each $x_i$,
the fitting can be rewritten as
\begin{align*}
    \hat y_i = \hat \mu(x_i) &= \sum_{j=1}^n\frac{\sum_{m=1}^M y_jI(x_j\in R_m)}{\sum_{k=1}^n I(x_k\in R_m)}I(x_i\in R_m)\\
    &=\sum_{j=1}^n\frac{\sum_{m=1}^M I(x_j\in R_m)I(x_i\in R_m)}{\sum_{k=1}^n I(x_j\in R_m)}y_j\triangleq \sum_{j=1}^n \bS_{ij}y_j\,,
\end{align*}
which yields $\hat\bfy = \bS\bfy\,.$
We have
\begin{align*}
\trace(\bS) = \sum_{i=1}^n \bS_{ii} &= \sum_{i=1}^n \sum_{m=1}^M\frac{I(x_i\in R_m)}{\sum_{k=1}^n I(x_k\in R_m)} = \sum_{m=1}^M\frac{\sum_{i=1}^n I(x_i\in R_m)}{\sum_{k=1}^n I(x_k\in R_m)} = M\,. 
\end{align*}
It follows that the \emph{modified search degrees of freedom} is $\msdf = \df(\hat\mu) - M$, \textred{where $\df(\hat\mu)$ is given by Equation~\eqref{eq:def_df}.}
On the other hand, the \emph{search degrees of freedom} might not be proper for the regression trees since the active variable set $\cA$ is not clear to define. And even we can identify the active set, it always relates the \emph{search degrees of freedom} to the full least squares on the active set instead of the regression tree itself.

The partition can be found by a greedy binary partition algorithm, followed by an optional pruning procedure \parencite{breimanClassificationRegressionTrees1994}. 
For simplicity, we skip the pruning procedure. The resulting tree will be complete, then the depth and the number of terminal nodes $M$ satisfy $M=2^\text{depth}$.
Table \ref{tab:df_regtree} shows the empirical degrees of freedom under different depths for simulated examples $p=1,5$ and $10$. Except for $M=1$, all $\hat\df$'s are much larger than the corresponding number of coefficients $M$. Similarly, we account for the surplus as the cost for \emph{searching} the partition variable and the associated cutpoint. \textcite{yeMeasuringCorrectingEffects1998} also did similar experiments which revealed a similar phenomenon for the regression tree.

\begin{table}
  \centering
    \caption[Empirical Degrees of Freedom of Regression Trees]{Empirical degrees of freedom of regression trees on simulated examples $p=1,5,10$ and $n=100$. The results are averaged over 10 simulations, with the standard error in parentheses.}
  \label{tab:df_regtree}
  \begin{tabular}{lcccc}
\toprule
\multirow{2}{*}{depth} & \multirow{2}{*}{$M$} & \multicolumn{3}{c}{$\hat\df$}\tabularnewline
\cmidrule{3-5}
& & $p=1$ & $p=5$ & $p=10$\tabularnewline
\midrule
0 & 1& 1.01 (0.05)& 1.02 (0.04)& 1.00 (0.05)\tabularnewline
1 & 2& 5.71 (0.08)& 8.71 (0.07)& 9.87 (0.10)\tabularnewline
2 & 4& 11.84 (0.20)& 18.38 (0.17)& 21.62 (0.22)\tabularnewline
3 & 8& 19.80 (0.18)& 30.14 (0.27)& 35.01 (0.35)\tabularnewline
4 & 16& 28.97 (0.29)& 42.06 (0.44)& 48.49 (0.35)\tabularnewline
\bottomrule
\end{tabular}

\end{table}

\subsection{Multiple Adaptive Regression Splines}\label{sec:mars}

Multiple Adaptive Regression Splines (MARS) is an adaptive procedure for regression proposed by \textcite{friedmanMultivariateAdaptiveRegression1991}. It is closely related to the tree-based method since it can be viewed as a generalization of stepwise linear regression of the tree regression method to improve the latter's performance \parencite{hastieElementsStatisticalLearning2009}. Specifically, with some minor changes, the MARS forward procedure will be the same as the tree-growing algorithm.

The model takes the form of an expansion in piecewise linear basis functions of the form $(x-t)_+$ and $(t-x)_+$. For each input $X_j$ and each observed value $x_{ij}$ of that input, construct the collection of basis functions,
$$
{\cal{D}}=\left\{(X_j-t)_+, (t-X_j)_+\right\}, \; t\in \{x_{ij}\}_{i=1}^n, j=1,\ldots,p\,.
$$
The model has the form $\mu(X) = \beta_0 +\sum_{m=1}^M\beta_mh_m(X)$,
where each $h_m(X)$ is a function in the collection $\cal{D}$, or a product of two or more such functions. If all $h_m(X)$ are restricted in $\cal D$, we call it an additive model (degree = 1), otherwise, we call it an interaction model (degree > 1) when there exist products of basis functions in $\cal{D}$.

MARS consists of a forward step and a backward step. The forward step adds the basis functions from the collection $\mathcal{D}$ into the model, either to be a new basis function or to multiply the existing function in the model. Similar to the pruning procedure in the tree-based methods, MARS also applies a backward deletion, and it uses the generalized cross-validation as the stop criterion,
$$
\GCV(M) = \frac{\sum_{i=1}^n(y_i-\hat \mu_M(x_i))^2}{(1-\tilde C(M)/n)^2}\,,
$$
where $\tilde C(M)$ is the effective number of parameters in the model, i.e., the degrees of freedom. It accounts both for the number of terms in the models, plus the number of parameters used in selecting the optimal positions of the knots. It is pre-determined before the model selection, so we call the \emph{nominal degrees of freedom}. 

\subsubsection{Search Cost and Nominal Degrees of Freedom}
\textcite{friedmanMultivariateAdaptiveRegression1991} proposed $\tilde C(M) = C(M) + c\cdot M$,
where $M$ is the number of nonconstant basis functions, $C(M)$ is the number of linearly independent basis functions, and the quantity $c$ represents the optimization cost for each basis function. He suggested that $c$ takes 2 for the additive model due to the expected decrease in the average-squared residual by adding a single knot to make a piecewise-linear model. If all basis functions (including the constant functions) are linearly independent, 
$\tilde C(M)$ coincides with the degrees of freedom formula in \textcite{friedmanFlexibleParsimoniousSmoothing1989} when $c$ is chosen to be 2.
\textcite{friedmanMultivariateAdaptiveRegression1991} also discussed the best value of $c$ in general cases. He mentioned that the best value for $c$ would depend on the number of basis functions, the number of samples, and the distribution of the covariates. Based on simulation studies, he suggested $c \in [2, 4]$, and recommended a ``fairly effective, if somewhat crude'' choice $c=3$. 

On the other hand, \textcite{hastieElementsStatisticalLearning2009} adopted a slightly different formula. To align the notations, let $r\triangleq C(M)$ be the number of linearly independent basis functions, and $K$ be the number of knots used in the forward procedure, then they wrote that
$$
\tilde C(M) = r + c\cdot K\,,
$$
where $c$ again takes 2 for additive models and 3 for interaction models. In other words, \textcite{hastieElementsStatisticalLearning2009} suggested $c$ cost for each knot instead of each basis function in \textcite{friedmanMultivariateAdaptiveRegression1991}. Note that each knot $t$ has two basis functions $(x-t)_+$ and $(t-x)_+$. In practice, \textcite{hastieMdaMixtureFlexible2022}'s R package \texttt{mda}\footnote{Line 256 in \texttt{mda\_0.5-3.tar.gz/src/dmarss.f}} and \textcite{milborrowEarthMultivariateAdaptive2021}'s R package \texttt{earth}\footnote{Line 1033-1034 in \texttt{earth\_5.3.1.tar.gz/src/earth.c}} determine the number of knots as $K = \frac{r-1}{2}$, and hence
\begin{equation}
\tilde C(M) = r + c\cdot \frac{r-1}{2}\,.
\label{eq:mars_cmr}
\end{equation}



\subsubsection{Correction for Self-consistency}

\begin{figure}
    \begin{subfigure}{0.33\textwidth}
    \includegraphics[width=\textwidth]{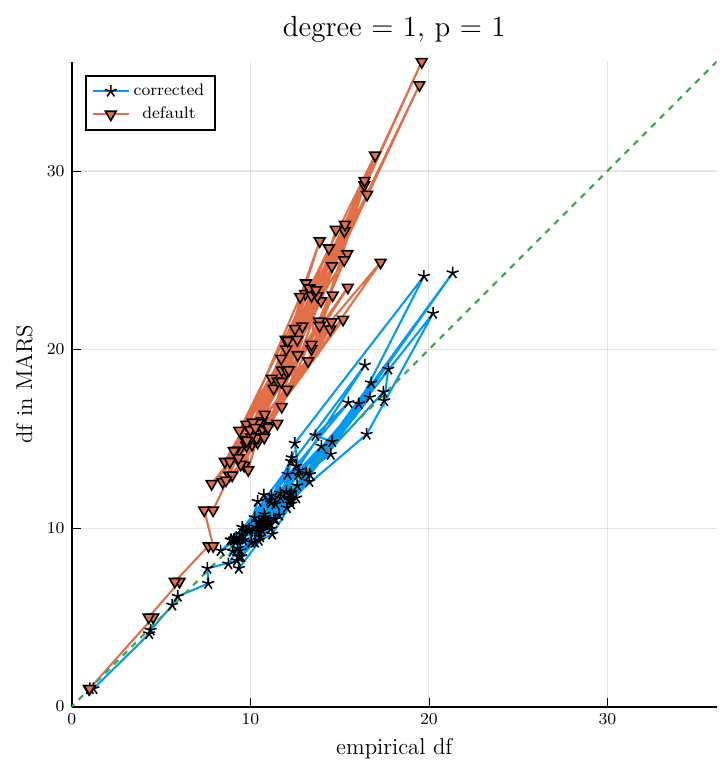}
    \end{subfigure}%
    \begin{subfigure}{0.33\textwidth}
    \includegraphics[width=\textwidth]{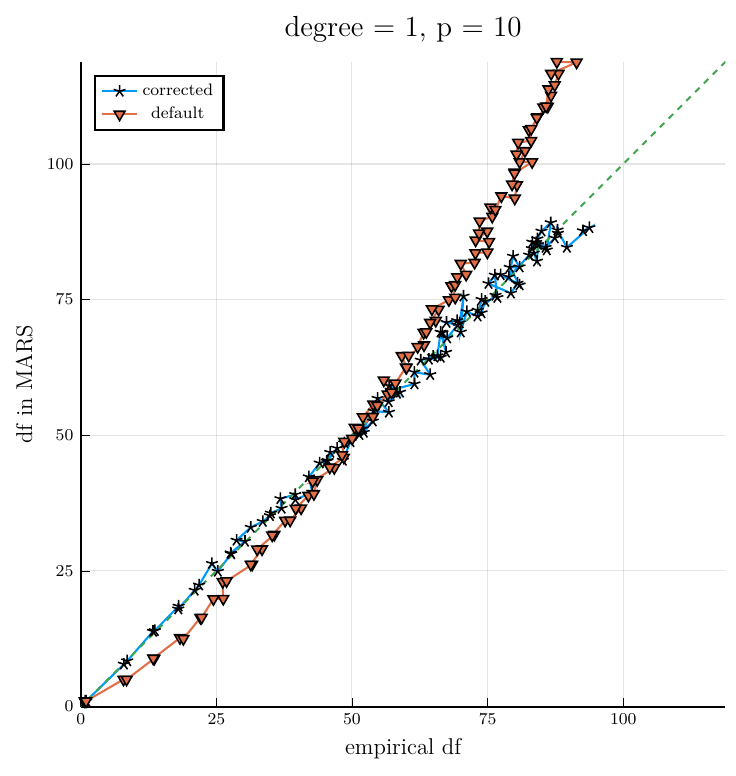}
    \end{subfigure}%
    \begin{subfigure}{0.33\textwidth}
    \includegraphics[width=\textwidth]{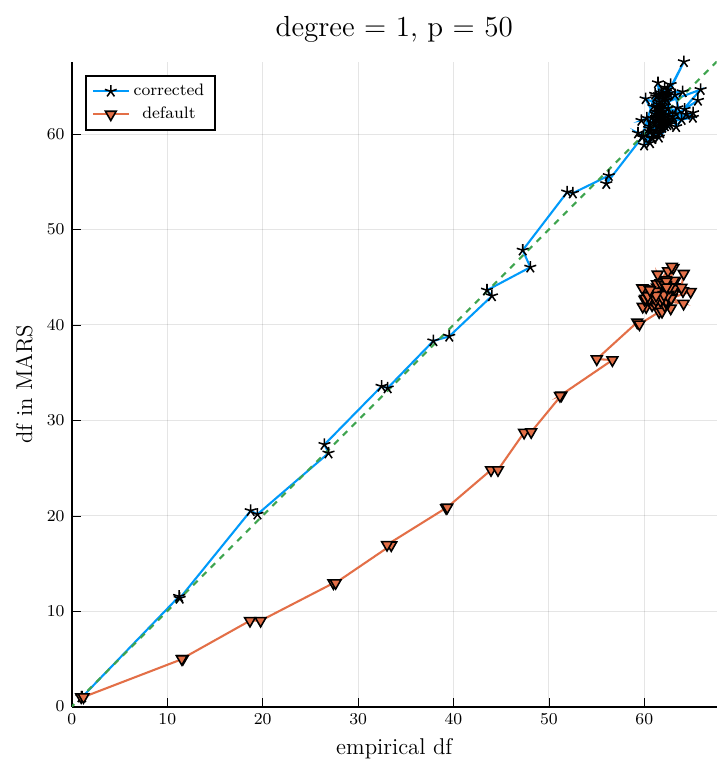}
    \end{subfigure}

    \begin{subfigure}{0.33\textwidth}
    \includegraphics[width=\textwidth]{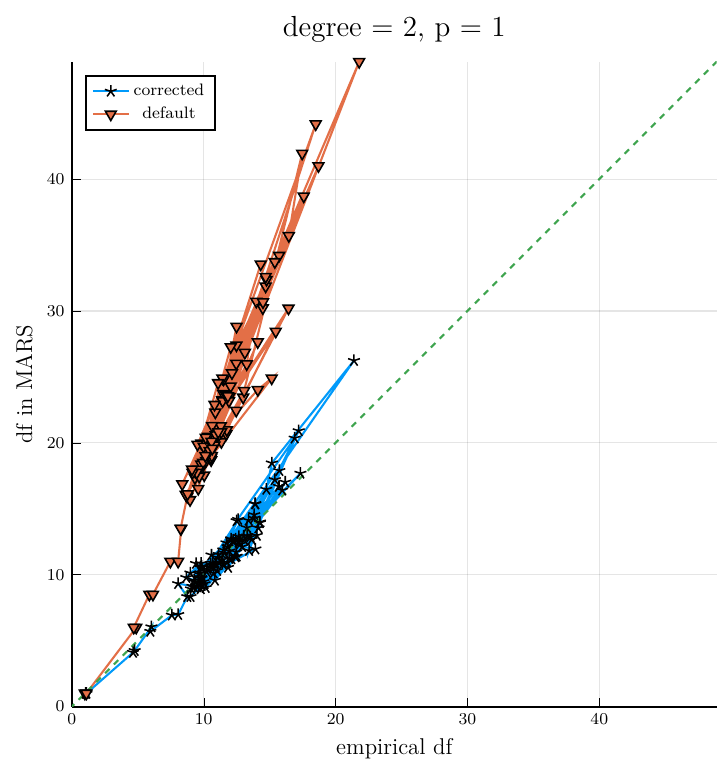}
    \end{subfigure}%
    \begin{subfigure}{0.33\textwidth}
    \includegraphics[width=\textwidth]{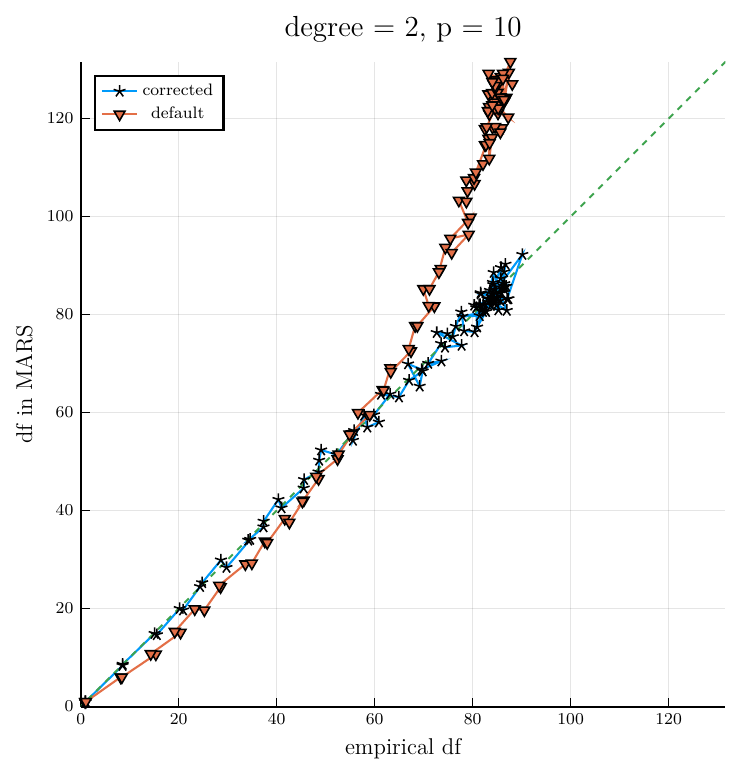}
    \end{subfigure}%
    \begin{subfigure}{0.33\textwidth}
    \includegraphics[width=\textwidth]{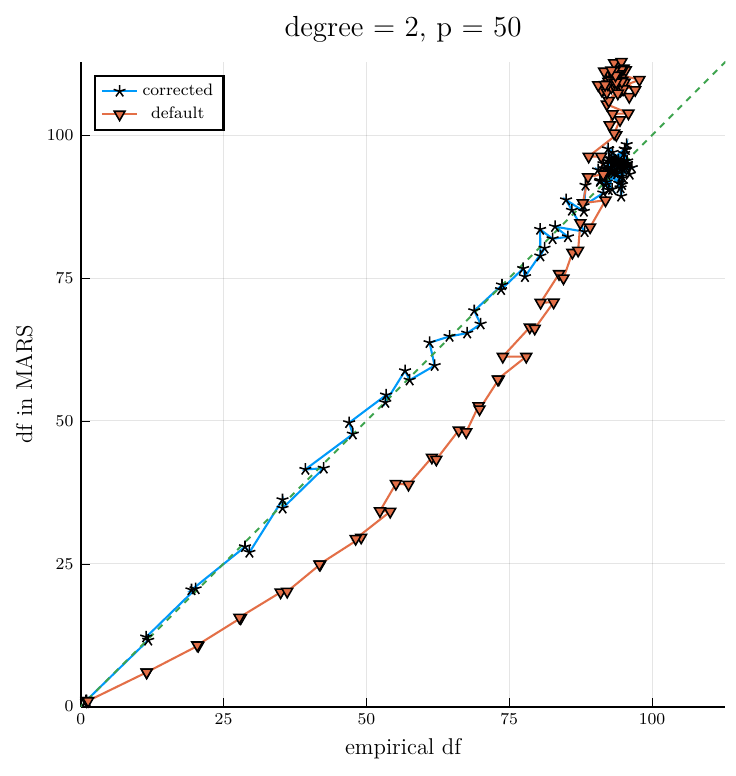}
    \end{subfigure}    
    \caption{Empirical degrees of freedom $\hat\df$ and MARS' degrees of freedom $\tilde C(M)$ with the default penalty factor and the corrected penalty factor in various scenarios indexed by the degree and the number of predictors $p$.}
    \label{fig:mars_df_vs_df}
\end{figure}

For MARS, the self-consistency does not hold since there is usually a gap between the \emph{actual degrees of freedom} $\hat\df$ approximated by Algorithm \ref{alg:dfhat} and the \emph{nominal degrees of freedom} $\tilde C(M)$, as shown in Figure \ref{fig:mars_df_vs_df}. 
Each point in Figure \ref{fig:mars_df_vs_df} represents a MARS fitting with $\tilde C(M)$ as $y$-coordinate and $\hat\df$ as $x$-coordinate. We vary the maximum number of knots (parameter \texttt{nk} in the function \texttt{earth::earth} from R package \texttt{earth}) in the forward procedure from 1 to 100, and for each \texttt{nk}, perform a MARS fitting, then connect the points along the parameter \texttt{nk}.
If the self-consistency property holds, the points will lie on the dashed line. 
The default setting denoted by triangle symbols refers to $c = 2$ for additive models (degree = 1) and 3 for interaction models (degree > 1), which is exactly the default setting in R package \texttt{earth}. The default setting is always away from the dashed line in either the additive model or the interaction model for different numbers of features.

To fulfill the self-consistency property, we allow $c$ to be a tuning parameter instead of the fixed values, 2 (for additive models) or 3 (for interaction models). 
 We propose Algorithm \ref{alg:correct_mars} to correct the penalty factor $c$. Figure \ref{fig:mars_df_vs_df} shows that the degrees of freedom corrected by Algorithm \ref{alg:correct_mars} lies on the dashed line, which indicates that the self-consistency property has been achieved using our algorithm. We also wrap up the algorithm in an R package \texttt{earth.dof.patch}\footnote{\url{https://github.com/szcf-weiya/earth.dof.patch}} for people who use MARS with its R package \texttt{earth}.
\begin{algorithm}[H]
    \caption{Correct Degrees of Freedom for MARS}
    \begin{algorithmic}[1]
        \STATE Calculate the empirical degrees of freedom $\hat \df$ for MARS (without pruning procedure) by Algorithm \ref{alg:dfhat}.
        \STATE Extract the nominal degrees of freedom $\tilde C(M)$ of MARS, and calculate $r = \frac{\tilde C(M) + c/2}{c/2+1}$ from Equation~\eqref{eq:mars_cmr}.
        \STATE Calculate corrected penalty factor by equaling the empirical $\hat \df$ to MARS's nominal degrees of freedom $\tilde C(M)$,
        $$
        c = \frac{2(\hat\df - r)_+}{r-1}\,.
        $$
    \end{algorithmic}
    \label{alg:correct_mars}
\end{algorithm}

\begin{remark}
    We restrict the MARS at step 1 without a pruning procedure to ensure a fixed value for $r$, otherwise both $r$ and $c$ in Equation \eqref{eq:mars_cmr} might vary. After obtaining a corrected penalty factor, one can freely choose to add a backward pruning procedure or not.
\end{remark}

To check whether the corrected degrees of freedom can improve the performance, we consider the tensor-product example in Section 9.4.2 of \textcite{hastieElementsStatisticalLearning2009},
\begin{equation}
Y = (X_1-1)_+ + (X_1-1)_+\cdot (X_2-0.8)_+ + 0.12\varepsilon\,,    
\label{eq:mars_sim_model}
\end{equation}
where the predictors $X_1,\ldots, X_p$ and errors $\varepsilon$ follow independent standard Gaussian distributions. \textred{There are $n=200$ observations.} Let $\mu(x)$ be the true mean of $Y$, and let
\begin{align*}
    \MSE_0 = \ave_{x\in \Test}(\bar y - \mu(x))^2\,,\quad \MSE = \ave_{x\in \Test}(\hat \mu(x) -\mu(x))^2\,,
\end{align*}
which represent the mean-square error of the constant model and the fitted MARS model, respectively. The proportional decrease in model error is
$$
R^2 = \frac{\MSE_0-\MSE}{\MSE_0}\,.
$$

\begin{figure*}
    \includegraphics[width=\textwidth]{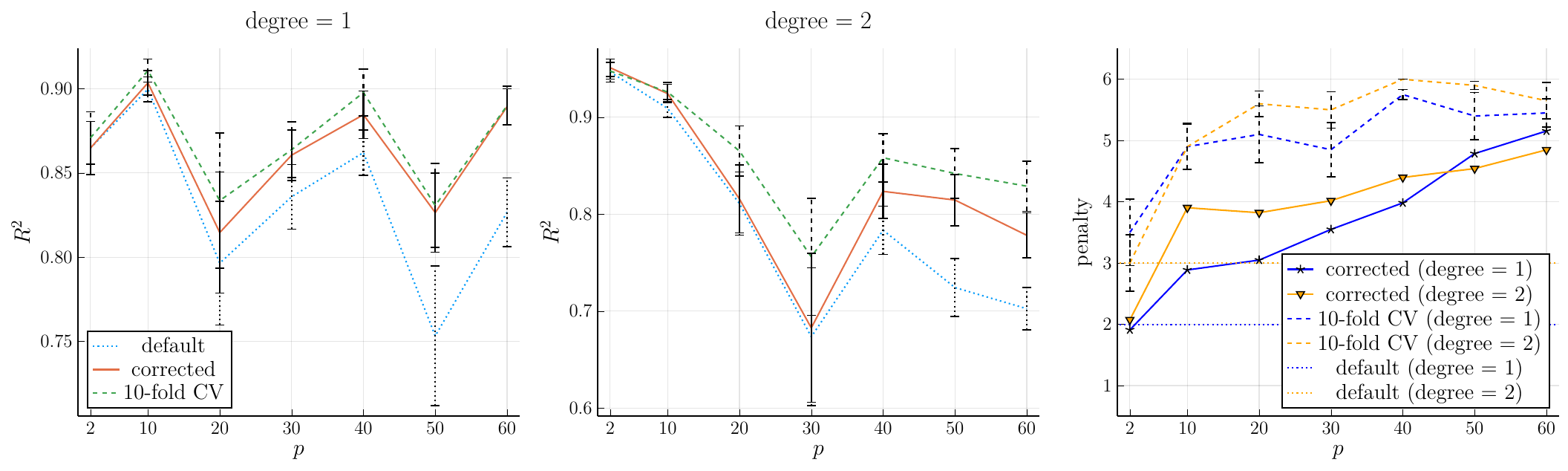}
    \caption{
    \textred{%
    Proportional decrease in model error $R^2$ when MARS with the default and corrected degrees of freedom are applied in different scenarios indexed by the number of predictors $p$. Two corrected degrees of freedom are considered: the proposed correction (``corrected'') for the self-consistency property, and the 10-fold cross-validation (``10-fold CV'').
    The left and middle panels display the average $R^2$ among 10 replications with error bars representing the standard errors along the number of predictors $p$ for degree = 1 and 2, respectively.
    The right panel shows the corrected penalty factors, where the error bar for the 10-fold CV is added since the penalty factor corrected by 10-fold CV varies for each repetition. }
    }
    \label{fig:mars_error_decrease}
\end{figure*}

Although the true model in Equation \eqref{eq:mars_sim_model} is generated with an interaction term, we consider the fitting with an additive model (degree = 1) in addition to fitting it with an interaction model (degree = 2). Figure \ref{fig:mars_error_decrease} shows the proportional decrease in model error $R^2$ when MARS with the default and corrected degrees of freedom are applied in different scenarios indexed by the number of predictors $p$. \textred{Besides the proposed correction for the self-consistency property, we also adopt a 10-fold cross-validation (CV) to select the optimal penalty factor that minimizes the CV error. The CV error curves can be found in the \supp.
In both additive and interaction model scenarios, both MARS with the proposed correction and MARS with the 10-fold CV can improve the MSE, especially when the number of predictors $p$ is large. The MARS with the 10-fold CV is even further better than the MARS with the proposed correction, but their gap is small as shown by the overlapped error bars.} The right panel shows the corrected penalty factor. \textred{Note that the 10-fold CV is conducted on each replication, and hence the penalty factor varies for each repetition.} When $p$ is small, the corrected penalty factor is smaller than the default penalty factor, and when $p$ is large, the corrected penalty factor is larger than the default one. The phenomenon is consistent with Figure \ref{fig:mars_df_vs_df}, where the \emph{nominal degrees of freedom} tends to be larger than the \emph{actual degrees of freedom} when $p = 1$, and smaller than the \emph{actual degrees of freedom} when $p$ is large. \textred{The correction by the 10-fold CV exhibits a similar increasing pattern to the proposed correction although the penalty factor by the 10-fold CV is usually larger. In summary, both the proposed correction and the correction by 10-fold CV show that the penalty factor should not be a constant across different $p$, and both can improve the performance compared to the default MARS. Note that the CV method is computationally extensive, and we need to conduct the CV step for each dataset, while the proposed corrected degrees of freedom can be viewed as an intrinsic property of the model itself, which is only calculated once and applicable to different datasets of the same dimension. Thus, our proposed correction is more efficient while achieving a comparable performance to the CV method.} 

\subsubsection{Real Data Applications}

We applied the original MARS and the corrected MARS to two datasets: the ozone data for a regression task and the spam email data for a classification task.\footnote{These results can be reproduced from the vignettes: \url{https://hohoweiya.xyz/earth.dof.patch/}.}

The ozone dataset is from the R package \texttt{earth}. It contains 330 observations on atmospheric ozone concentration, along with 9 meteorological variables, collected in the Los Angeles Basin in 1976. We used the MARS and the corrected MARS to predict the ozone concentration using other variables. To evaluate the prediction performance, we calculated the 5-fold cross-validation error, as shown in Table~\ref{tab:ozone}. The corrected MARS can achieve slightly smaller errors, but we also note that the performances are quite comparable since the mean differences are within the standard errors.

\begin{table}[H]
    \centering
    \caption{The 5-fold cross-validation error, with standard error in the parentheses, on the ozone data by the MARS and the corrected MARS.}
    \label{tab:ozone}
    \begin{tabular}{cccc}
    \toprule
        & MARS & corrected penalty factor & corrected MARS\\
        \midrule
      degree = 1   & 16.66 (2.29) & 2 $\rightarrow$ 2.99 & \textbf{16.61} (2.26) \\
      degree = 2   & 17.33 (2.34) & 3 $\rightarrow$ 3.92 & \textbf{17.20} (2.42)\\
      \bottomrule
    \end{tabular}
\end{table}

The spam email data \parencite{misc_spambase_94} consists of information from 4601 email messages, in a study to screen junk email (i.e., spam). The response variable is binary, with values \texttt{email} or \texttt{spam}, and there are 57 predictors. We split the dataset into a training set ($n = 3065$) and a test set ($n = 1536$) as in \textcite{hastieElementsStatisticalLearning2009}. The overall test error rate for the MARS and the corrected MARS are summarized in Table~\ref{tab:spam}, from which we can see that the corrected MARS can always achieve better performance regardless of the degree of interactions.

\begin{table}[H]
    \centering
    \caption{Test error rate for the MARS and the corrected MARS fit to the spam email data.}
    \label{tab:spam}
    \begin{tabular}{cccc}
    \toprule
         & MARS & corrected penalty factor & corrected MARS  \\
         \midrule
       degree = 1  & 0.0846 & 2 $\rightarrow$ 6.25 & \textbf{0.0814}\\
       degree = 2  & 0.0684 & 3 $\rightarrow$ 7.94 & \textbf{0.0658}\\
     \bottomrule
    \end{tabular}
\end{table}






\section{Discussions}\label{sec:df_future_work}

Through a number of model fitting procedures, we have shown that the degrees of freedom usually does not equal the number of free parameters. For adaptive approaches such as regression trees and best subset regressions, the degrees of freedom is larger than the number of the free parameters, and the excess amount is referred to as the \emph{search degrees of freedom}; for regularized methods such as ridge regressions and splines, the degrees of freedom would be smaller than the number of free parameters. We extend the definition and propose the \emph{modified search degrees of freedom}, which can account for the search cost of a linear operator. Remarkably, the degrees of freedom of the lasso is exactly the number of selected coefficients, but we take another perspective and find that the lasso also exhibits a nonzero search cost. The \emph{modified search degrees of freedom} also works for procedures with augmented spaces, such as splines methods, tree-based methods, and MARS. 
We also investigate the gap between the \emph{nominal degrees of freedom} and the \emph{actual degrees of freedom} when the degrees of freedom serve as a parameter in model selection. We define the \emph{self-consistency} property when there is no gap between these two degrees of freedom. For MARS, which violates the self-consistency property, we propose a correction procedure to fulfill the self-consistency property. It turns out that the corrected approach can significantly improve the fitting performance, as evidenced in simulations and real data applications.

Despite our efforts to improve the understanding of the degrees of freedom by developing the search cost and self-consistency concepts, here are some limitations that need future development.
\begin{itemize}
    \item \textred{The modified search degrees of freedom offers a broader perspective on the search cost for a wider range of procedures than \textcite{tibshiraniDegreesFreedomModel2015}'s search degrees of freedom. However, our understanding of its practical benefits is still in the early stages. Here are two potential applications. First, we can diagnose the relationship between the search cost and other parameters, such as the number of predictors. For example, it reveals that the search cost increases with the number of predictors, as demonstrated in Table~\ref{tab:df_regtree} for regression trees and Figure~\ref{fig:mars_error_decrease} for MARS. Another potential application is to compare the complexity of different models by quantifying the cost of searching in the model space while ignoring the cost of fitting a fixed model. This focus on the search aspect can provide insights into the efficiency and feasibility of different model selection procedures.}
  \item The general definition in Equation \eqref{eq:def_df} assumes homogeneous variance, but practically there are many heterogeneous cases. There is a need to discuss the generalization to heterogeneous situations.
  \item The phenomenon that the degrees of freedom might be larger than the number of coefficients can be easily observed in simulations, but there is little theoretical support and discussion in the literature. It would be more helpful to derive some closed form for the degrees of freedom, at least in some special cases, such as the best subset regression (Section \ref{sec:subset}) in the orthogonal setting in \textcite{tibshiraniDegreesFreedomModel2015}. 
  \item As suggested from the definition in Equation \eqref{eq:def_df}, the degrees of freedom focuses on the in-sample prediction $\hat\mu_i$ instead of out-of-sample prediction. \textcite{luanPredictiveModelDegrees2021} extended the definition to out-of-sample prediction, and termed as \emph{predictive} degrees of freedom. The authors claimed that it could help explain the ``double descent'' phenomenon in over-parametrized interpolating models (e.g., \textcite{zhangUnderstandingDeepLearning2021} and \textcite{hastieSurprisesHighdimensionalRidgeless2022}). Currently, their predictive degrees of freedom is only discussed for linear regressions. It would be interesting to check more connections with other models, such as MARS in Section \ref{sec:mars}.
\end{itemize}

\section*{Acknowledgement}
The main results in this article are developed from Lijun Wang's Ph.D. thesis when he was at the Chinese University of Hong Kong under the supervision of Xiandan Fan. Lijun Wang was supported by the Hong Kong Ph.D. Fellowship
Scheme from the University Grant Committee. Xiaodan Fan was supported by two grants from the Research Grants Council (14303819, C4012-20E) of the Hong Kong SAR, China.

\section*{Disclosure Statement}

The authors declare no competing interests.

\section*{Supplementary Material}

The \supp{} contains technical proofs for theoretical results and additional simulation studies.

\spacingset{1.0}
\bibliographystyle{biometrika}
\printbibliography

\end{document}